\documentclass{PoS}

\title{Jet Formation in MHD Simulations}

\ShortTitle{Jet Formation in MHD Simulations}

\author{\speaker{P. Chris Fragile}%
         \thanks{Special thanks to the SOC and LOC, NSF, SCSGC, and SC EPSCoR.}\\
        College of Charleston\\
        E-mail: \email{fragilep@cofc.edu}}

\abstract{In this talk I review the current status of jet formation in direct numerical simulations of black-hole accretion disks and magnetospheres. I address the following critical questions: What constitutes the jet? What is the launching mechanism? Where is the launching point of the jet? What is the Lorentz factor? What is the opening angle? How is the jet collimated? Just as importantly, I also discuss how dependent the answers to the above questions are on factors such as the initial conditions of the simulation. I end by discussing possible future directions for this research.}

\FullConference{VII Microquasar Workshop: Microquasars and Beyond\\
		 September 1 - 5, 2008\\
		 Foca, Izmir, Turkey}

\begin{document}

Let me begin by saying a few words about what I will not cover in this review. I plan to stick to a narrow interpretation of the title given to me by the Scientific Organizing Committee. Firstly, I will be dealing mostly with direct numerical simulations; I will not present any observational results and will only briefly mention one or two well-known theories. Secondly, I will specifically focus on numerical simulations of the \emph{formation} of jets, ignoring the large body of literature on the evolution and stability of jets and their interactions with their environments, where the jets are taken as an input condition. I will instead carefully focus on simulations that include complete treatments of the accretion disks and the region right around the event horizon of the black hole, where the jets are actually launched.

A convenient place to start this review is to look at where this field was at the time of the 6th Microquasar Workshop in 2006. That also happened to be the year that two of the main papers on numerical simulations of black-hole jet formation were published, one by McKinney \cite{mckinney_06} and the other by Hawley \cite{hawley_06}. After summarizing their work, I will discuss some of the important things that have been learned about jet formation in MHD simulations in the two years since then. I will finish by discussing some possible future directions for this research, some of which is already under way.

\section{What Constitutes the Jet?}
\begin{figure}
\begin{center}
\includegraphics[width=0.4\textwidth]{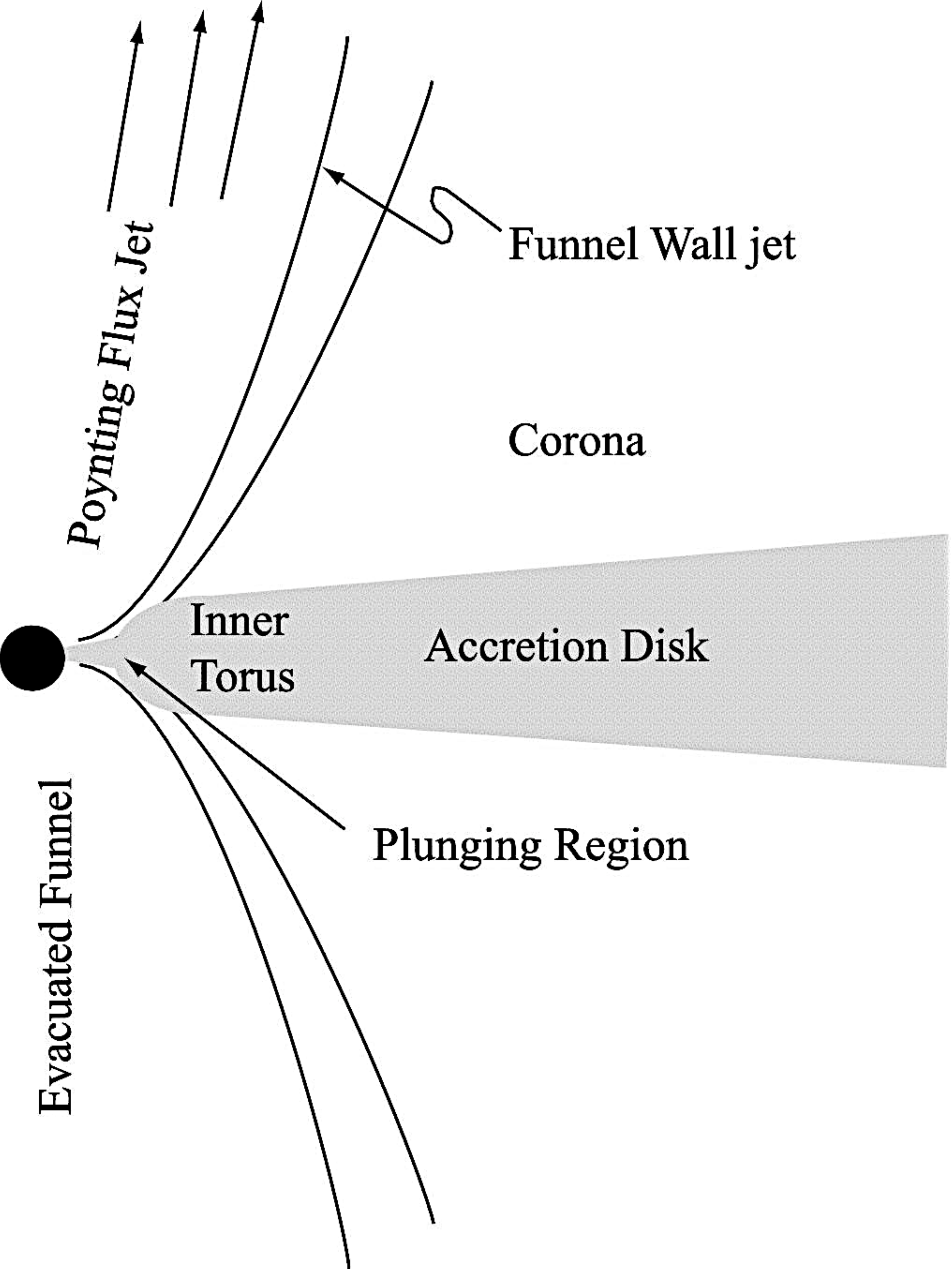}
\includegraphics[width=0.4\textwidth]{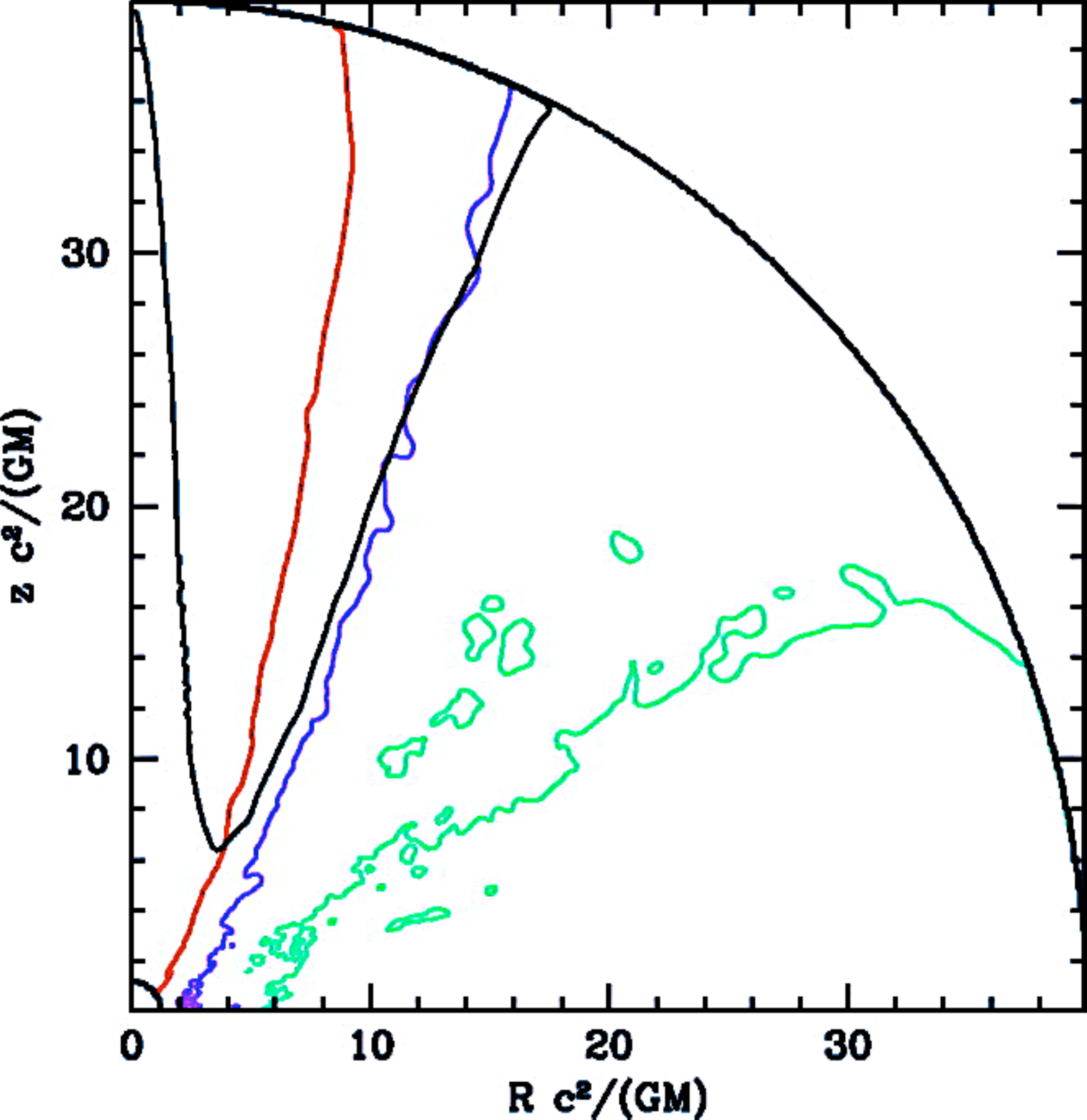}
\end{center}
\caption{{\em Left panel:} Schematic illustration of the main dynamical features seen in the accretion disk simulations of Hawley \cite{hawley_06}. The jet is defined as those regions that are unbound and have positive radial momentum. {\em Right panel:} A more quantitative delineation of the different regions from the simulations of McKinney \cite{mckinney_04}. The material inside the {\em green} contour, corresponding to the main disk body, is weakly magnetized ($\beta=P_{gas}/P_{mag}\ge 3$). The material inside the {\em violet} contour, corresponding to a magnetized corona, is near equipartition ($1 \le \beta \le 3$). All of the material inside the {\em black} contour is unbound ($u_t \le -1$) and would be called ``jet'' material. Inside the {\em red} contour, the magnetic field energy density dominates even over the rest mass density of the fluid ($B^2/\rho \ge 1$).}
\label{fig1}
\end{figure}

Let me first be clear about what is meant by the term ``jets'' when referring to numerical simulations, as it is not obvious at this point how they are related to the jets observed in Nature. In the simulations of McKinney \cite{mckinney_04} and Hawley \cite{hawley_06}, the jets are identified as unbound ($u_t \le -1$), outflowing ($u^r > 0$) material, where $u^\mu$ is the 4-velocity of the fluid. It is actually fairly remarkable how similar the results of such numerical simulations are, despite a number of differences in the particulars of the codes that are used. A quick comparison shows that the numerical simulations of both groups produce qualitatively and quantitatively similar final states, which include prominent jet-like features. The various features identified in these simulations are illustrated in Figure \ref{fig1}.

For most of the rest of this review I will distinguish between the two different jet components shown in Figure \ref{fig1}: the Poynting-flux jet and the matter-dominated or funnel-wall jet. Despite the strong similarities in their results, McKinney and Hawley chose to focus on different aspects of their results. Most of McKinney's analysis is concentrated on the Poynting-flux jet, whereas Hawley generally discusses the matter-dominated jet in more detail. This will make it easy for me to keep the two separated in this review.

\section{The Launching Mechanism}

\subsection{Poynting-Flux Jet}
\begin{figure}
\begin{center}
\includegraphics[width=0.4\textwidth]{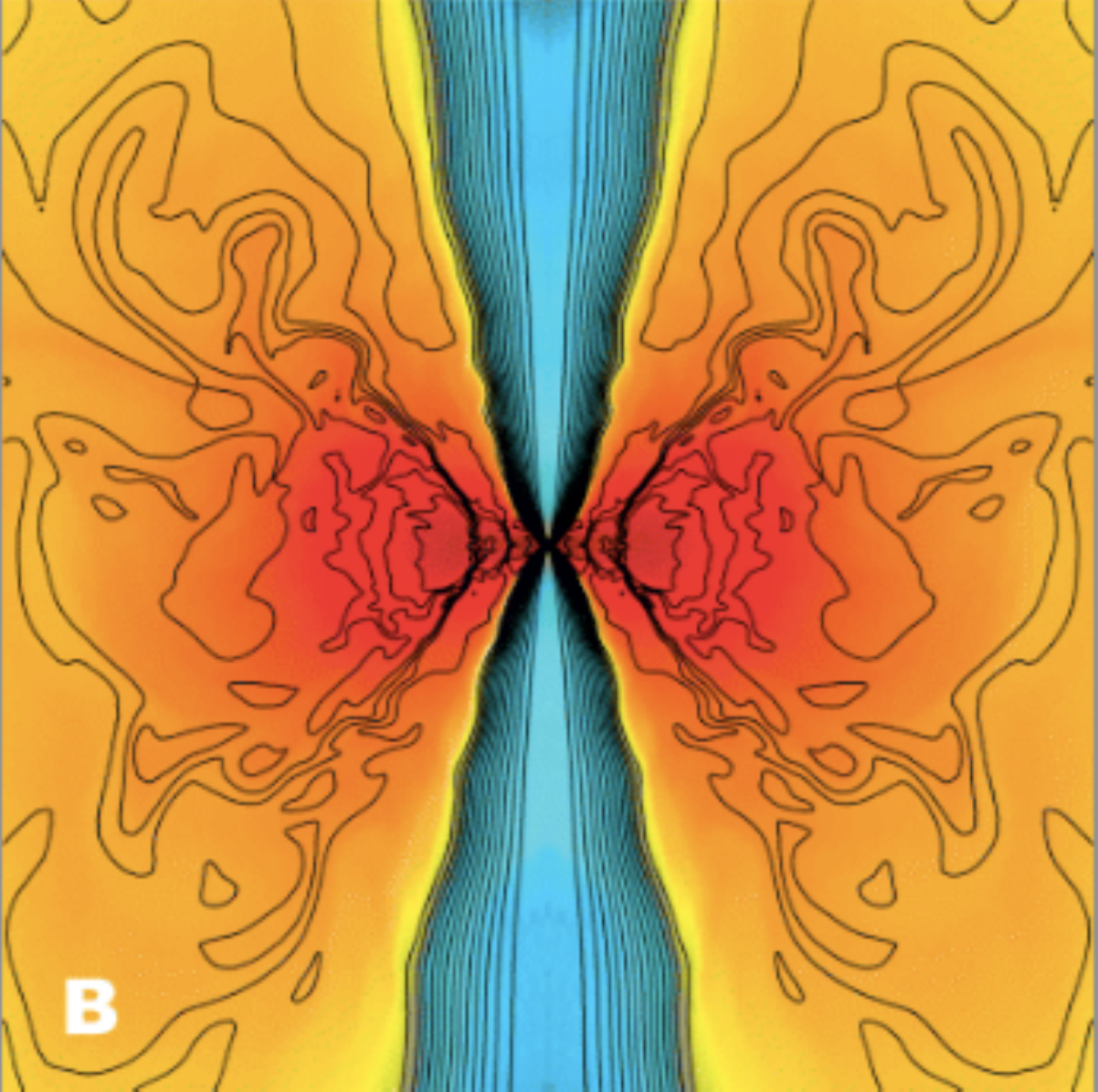}
\end{center}
\caption{Magnetic field ({\em black lines}) overlaid on a pseudocolor plot of the log of density from the simulations of McKinney \cite{mckinney_06}}.
\label{fig2}
\end{figure}

Figure \ref{fig2} shows magnetic field lines overlaid on a pseudocolor plot of density from the simulations of McKinney \cite{mckinney_06}. The important things to note about that figure are that, although the magnetic field lines appear tangled and chaotic inside the disk, as expected from the action of the magneto-rotational instability (MRI), inside the nearly evacuated funnel region the field lines are well ordered all the way down to the event horizon of the black hole. In fact, the magnetic field lines have the characteristic appearance of the split-monopole configuration described by Blandford \& Znajek \cite{blandford_77}, suggesting a possible launching mechanism. 

It is worth emphasizing at this point that the initial configuration for McKinney's simulation did {\em not} include any large-scale magnetic fields. Instead the simulation started with only a weak dipole magnetic field in the form of poloidal loops contained entirely within the disk (see Figure \ref{fig8} below for an illustration). The fields that subsequently fill the evacuated funnel and power the Poynting-flux jet arise solely from these initial seed fields and the evolution of the disk. This is important because it means that powerful jets do {\em not} require large-scale, ordered magnetic fields be fed in from large radii through the disk.

\begin{figure}
\begin{center}
\includegraphics[width=0.4\textwidth]{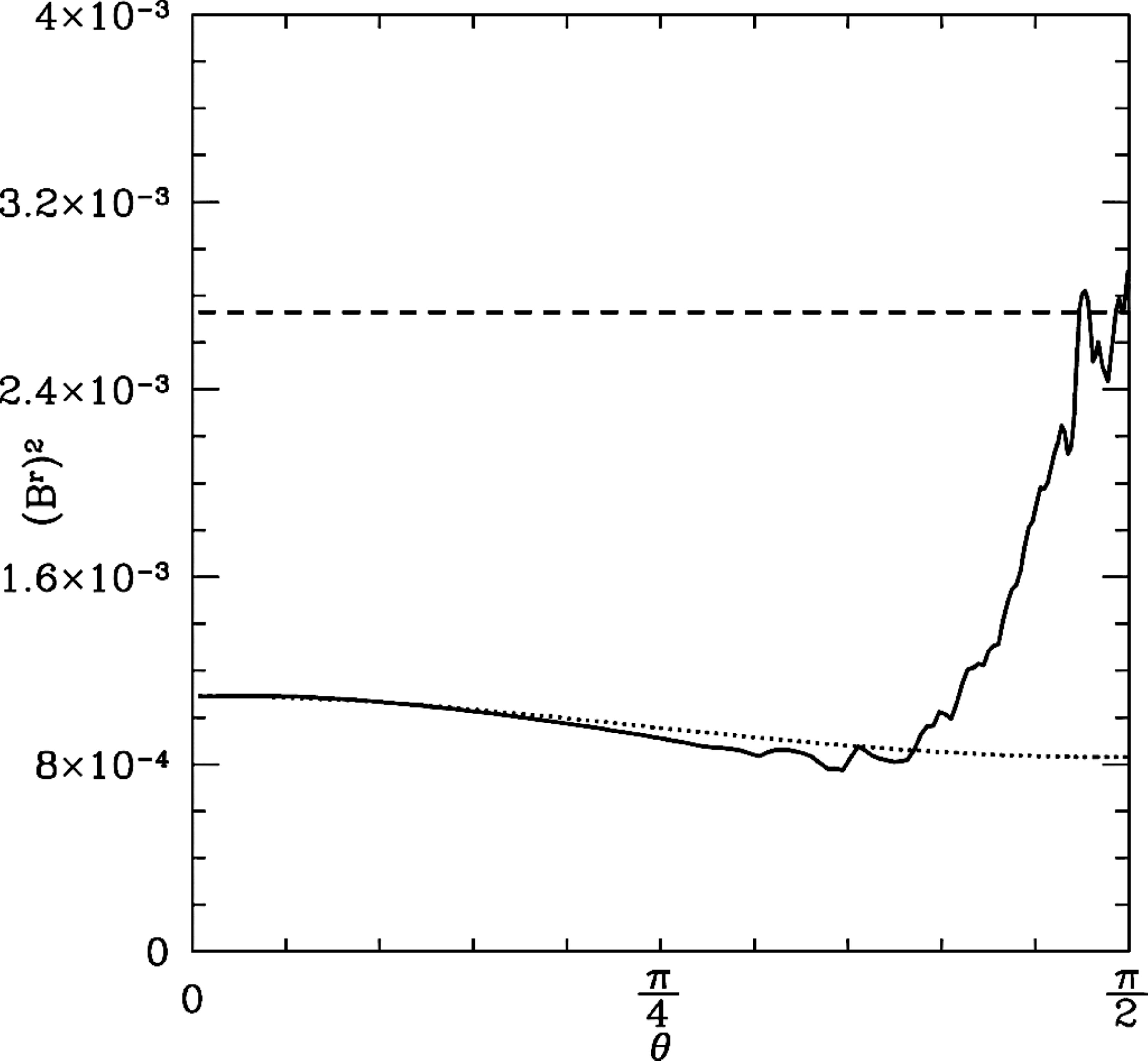}
\includegraphics[width=0.4\textwidth]{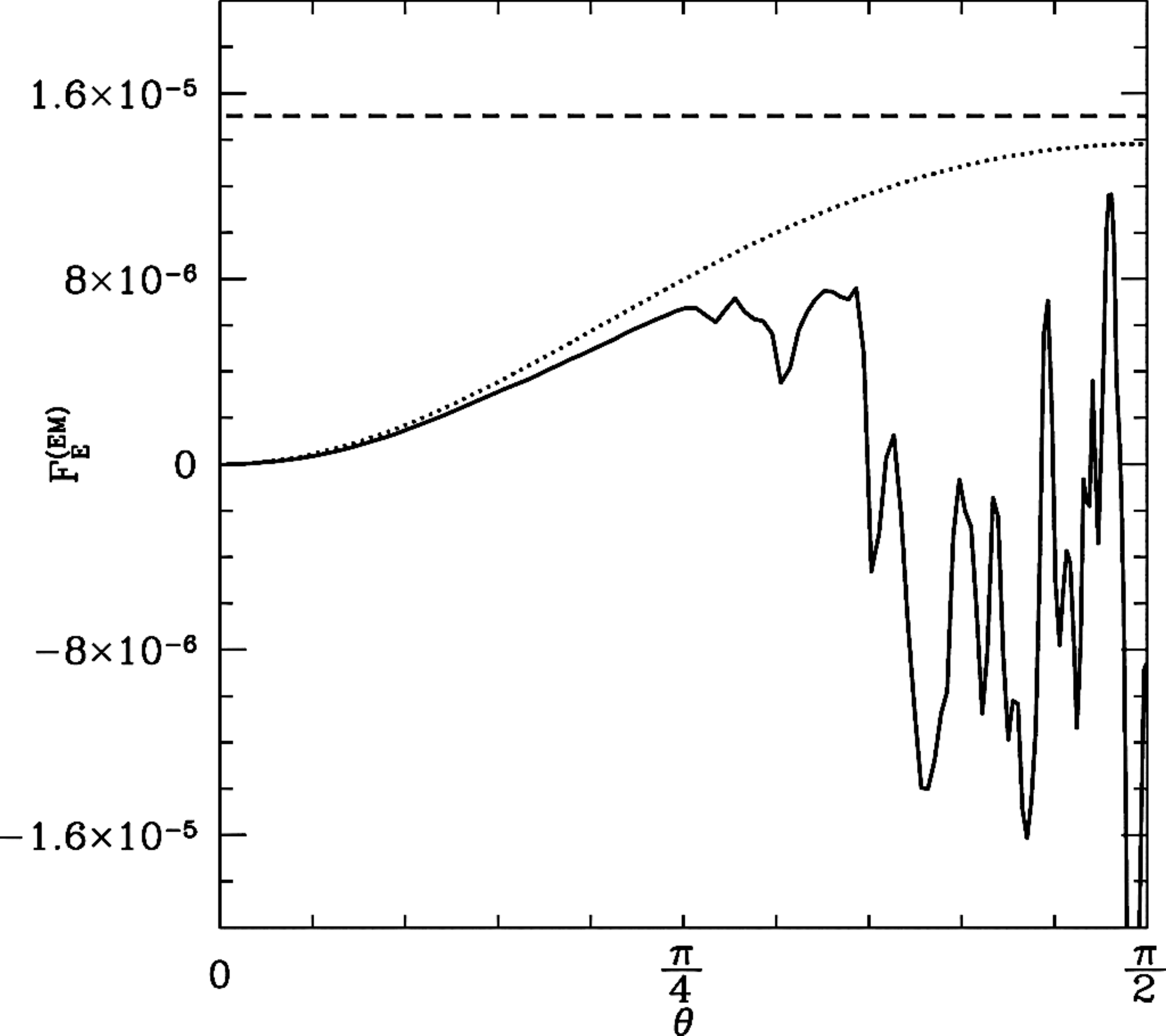}
\end{center}
\caption{{\em Left panel:} Square of radial field on the horizon from the time- and hemisphere-averaged simulation data of \cite{mckinney_04} ({\em solid line}). The {\em dotted line} shows the Blandford-Znajek \cite{blandford_77} perturbed monopole solution with the field strength normalized to the numerical solution at the pole. {\em Right panel:} Electromagnetic energy flux  on the horizon from the time- and hemisphere-averaged data of \cite{mckinney_04} ({\em solid line}). The dotted line shows Blandford-Znajek's spun-up monopole solution. }
\label{fig3}
\end{figure}

Although the field configuration in Figure \ref{fig2} is strongly suggestive that the Poynting-flux jets are powered by the Blandford-Znajek process, it is not conclusive. Fortunately the case is made much stronger by the careful quantitative comparisons McKinney \& Gammie made between their simulations and the predictions of the Blandford-Znajek model \cite{mckinney_04}. Examples of the results of such comparisons are shown in Figure \ref{fig3}. Because the Blandford-Znajek model does not include a realistic accretion disk, one must be careful when comparing it to numerical simulations of disks, such as the ones being considered here. It is only realistic to expect the Blandford-Znajek solution to apply in regions where the magnetic field is dominant, as in the nearly evacuated funnel region. From Figure \ref{fig3} we see that in this region ($\theta < \pi/4$) the numerical simulation agrees quite well with the model. This would seem to confirm that the Poynting-flux jets are indeed powered by the Blandford-Znajek mechanism (magnetic field lines threading the ergosphere of the black hole tap the rotational energy of the black hole and transport that energy to large radii).

\subsection{Matter-Dominated Jet}
\begin{figure}
\begin{center}
\includegraphics[width=0.4\textwidth]{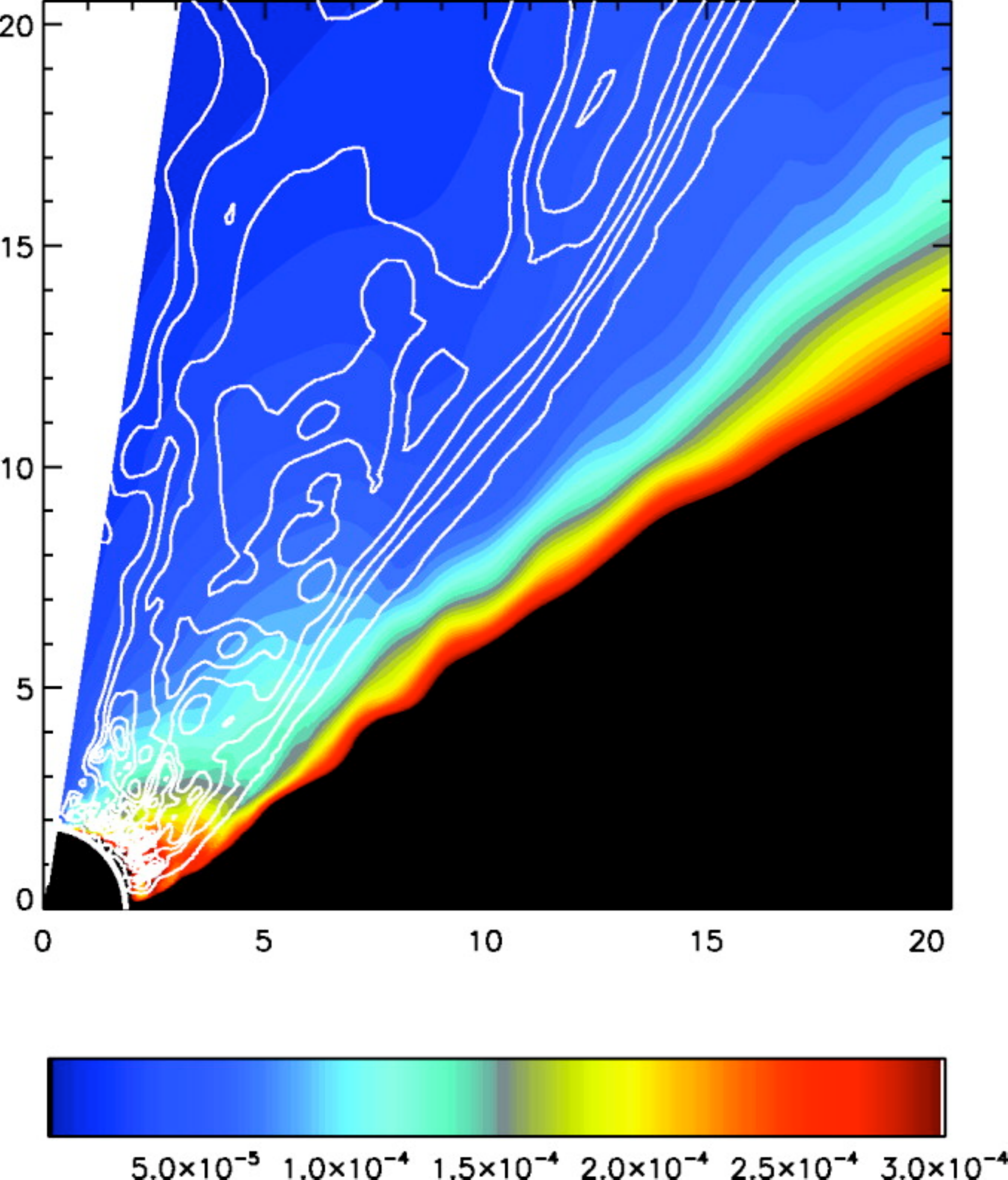}
\end{center}
\caption{Color contours of the azimuthal average of the total pressure (gas plus magnetic). Radial momentum flux in the outflow is shown in white contours. Both sets of contours are linear scales and come from the simulations of Hawley \& Krolik \cite{hawley_06}. }
\label{fig4}
\end{figure}

The matter-dominated jet has a much different launching mechanism. In some ways it is even easier to understand, particularly if one can understand Figure \ref{fig4}, taken from the simulations of Hawley \& Krolik \cite{hawley_06}. Notice in the figure that there is a strong pressure gradient along the coronal wall that points oblique to the momentum-flux contours. Breaking this pressure gradient into a component perpendicular to the momentum-flux contours and a component parallel, we can easily see that the pressure gradient serves two roles: 1) the perpendicular component pushes matter into the jet (mass loading); and 2) the parallel component accelerates matter along the jet. These strong pressure gradients line the entire funnel wall, so there is a very extended mass loading and acceleration region.

\section{The Launching Point}
\begin{figure}
\begin{center}
\includegraphics[width=0.4\textwidth]{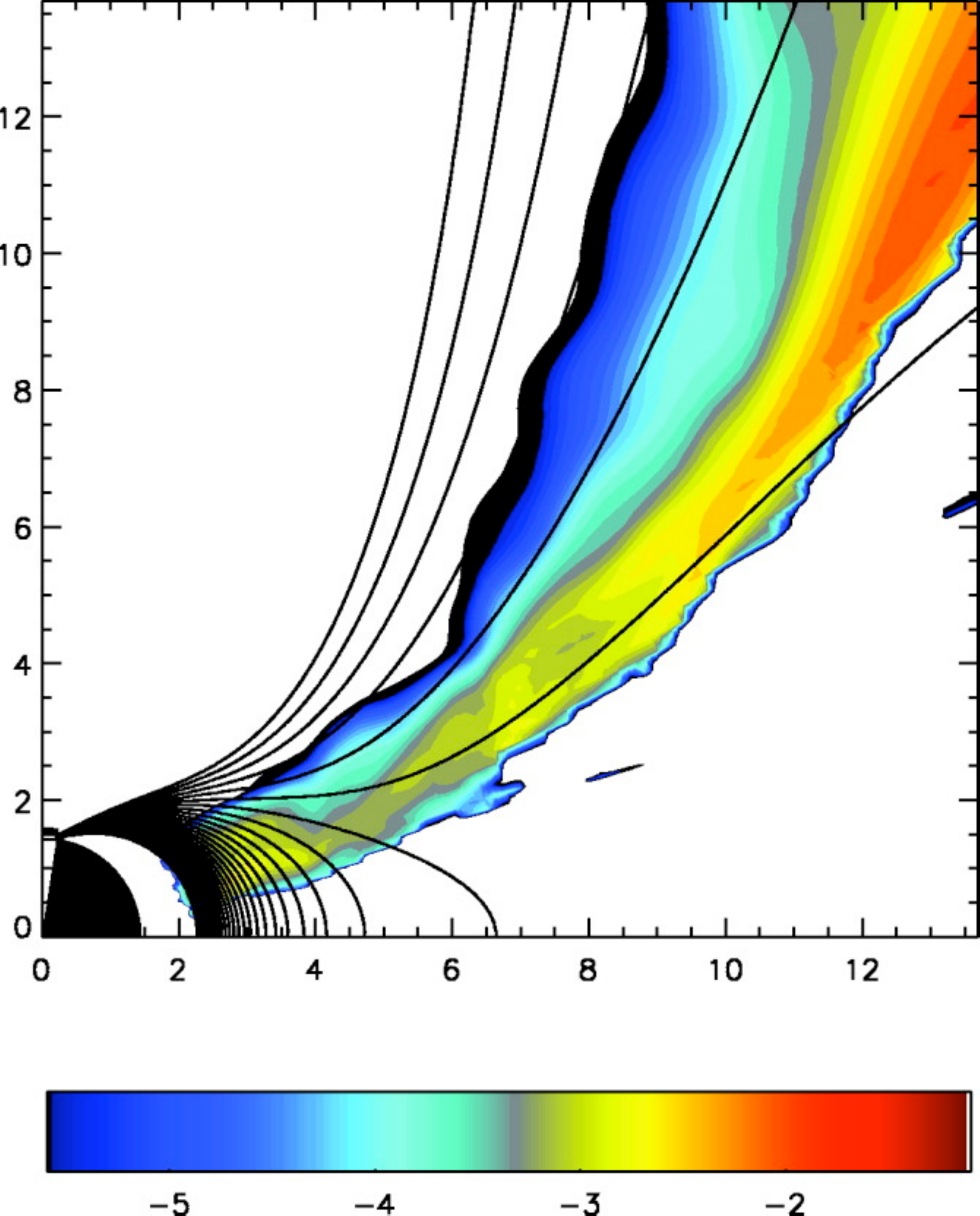}
\end{center}
\caption{Mass outflow and effective potential contours in simulation KDR from \cite{hawley_06}. Here, the color contours are the logarithm of the mass outflow rate, while the line contours are linear in the potential.}
\label{fig5}
\end{figure}

We can confirm that the acceleration of the matter-dominated jet continues all along the funnel wall by looking at the plot of mass outflow rate shown in color in Figure \ref{fig5}. Note that the mass outflow rate continues to increase out to the outer radial extent of this figure ($r\sim 15r_G$). Figure \ref{fig5} also shows the equipotential surfaces ({\em black lines}) of the flow. Motion of gas away from the black hole across the equipotential lines inside $r\lesssim 6 r_G$, means that the jet material is climbing out of the potential well of the black hole. As shown in Figure \ref{fig4}, the driving force behind this acceleration is the strong gradient in total pressure near the hole. Another important point illustrated in Figure \ref{fig5} is that outside of $r \gtrsim 6r_G$, the matter-dominated jet flows {\em parallel} to the equipotential surfaces, meaning that the acceleration must not be attributable to centrifugal forces. This strongly rules out a launching mechanism such as the Blandford-Payne wind model \cite{blandford_82}.

\section{Lorentz Factor}
Now we turn to consider characteristics of the jets in these simulations that might be compared directly with observations, such as the Lorentz factor and opening angle. For the matter-dominated jet, measuring the Lorentz factor from numerical simulations is a rather straightforward operation, and a range of values are quoted in the literature. Typical values are $\Gamma \sim 1.4-1.5$ \cite{mckinney_04,komissarov_05}.

However, for the Poynting-flux dominated jet, measuring a Lorentz factor from numerical simulations like those of McKinney or Hawley is a difficult proposition. This is because the evacuated funnel region occupied by the Poynting-flux jet can not be handled properly with the numerical techniques that are employed. This is because the evacuated funnel should really be evacuated (void of any matter) near what is called the stagnation point. The stagnation point, quite simply, is a point along each field line, inside of which matter is falling toward the black hole and outside of which the matter is moving off to infinity. In ideal MHD (as treated in the numerical codes of Hawley and McKinney), matter is not allowed to move across field lines, so it is impossible to replenish matter as it is drained away from the stagnation point. Eventually, the density there must go to zero. However, in the numerical codes used in the simulations I've been describing, the gas density can never be allowed to go to zero without causing the code to crash. Therefore, matter must artificially be added to the simulation in the vicinity of the stagnation point and anywhere else the density falls below some threshold. Because the mass content of the jet is a main factor in determining the Lorentz factor, this artificial mass-loading can have a profound impact on the Lorentz factor measured in the numerical simulations. A realistic Lorentz factor could only be obtained if there were a valid model for the mass loading of the jet.

\section{Opening Angle}
\begin{figure}
\begin{center}
\includegraphics[width=0.5\textwidth]{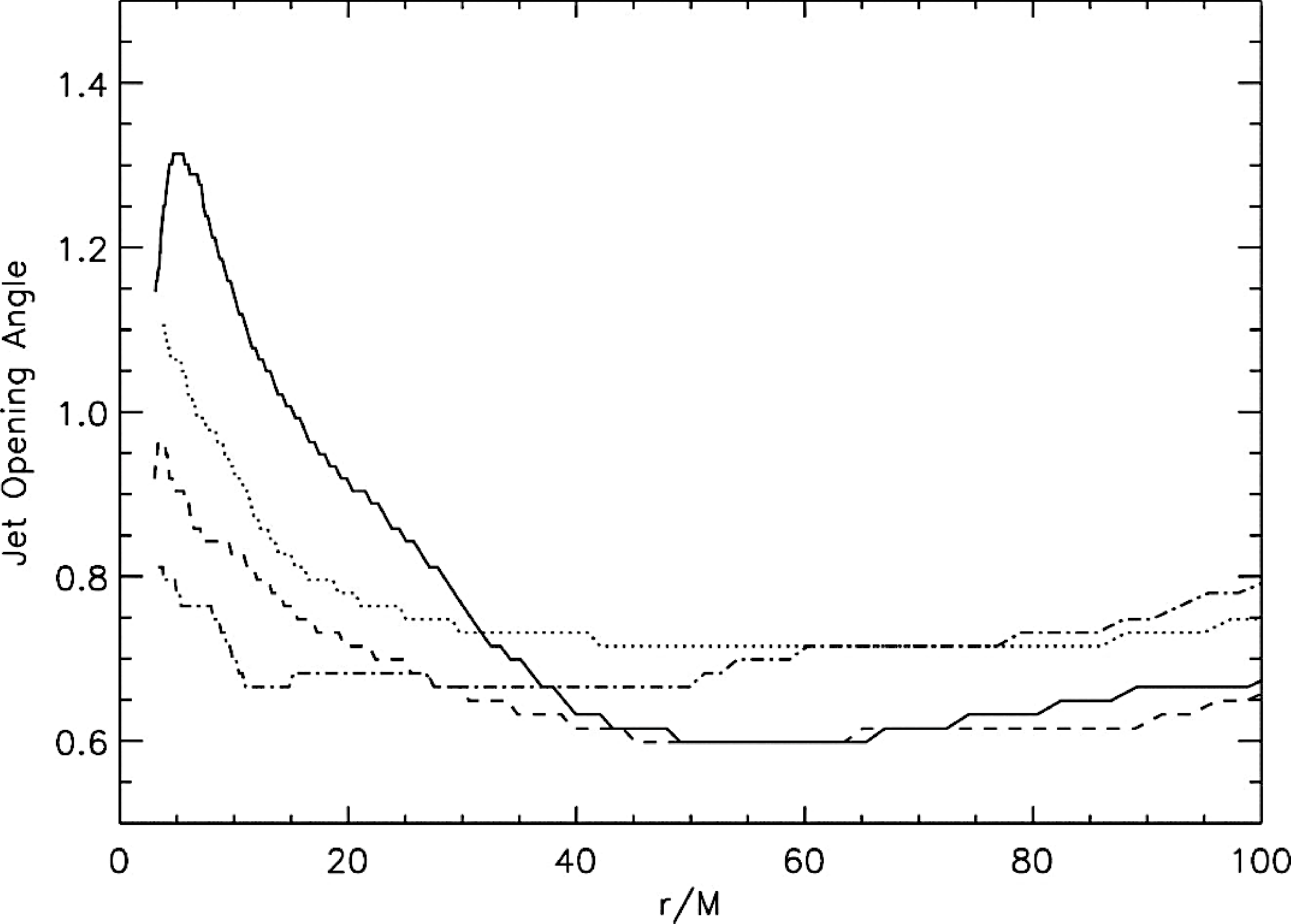}
\end{center}
\caption{Opening angle of the jet in radians, as defined by the angular location of the boundary between bound and unbound outflow, from the simulations of Hawley \& Krolik \cite{hawley_06}. Four models are shown: the solid line is $a/M =-0.9$, the dotted line is $a/M = 0.5$, the dashed line is $a/M = 0.9$, and the dot-dashed line is $a/M = 0.99$.}
\label{fig6}
\end{figure}

Along with being relatively slow, the matter-dominated jets also have fairly large opening angles, as shown in Figure \ref{fig6} from Hawley \& Krolik \cite{hawley_06}. The initial opening angle at the base of the jet is in the range $50^\circ \lesssim \theta \lesssim 60^\circ$. As the matter-dominated jet moves outward, some collimation is provided by the coronal walls, so that at $r\sim 40r_G$ the opening angle is down to $40^\circ$. Still, this seems rather too large to agree with most observed jets. However, it may be that there are mechanisms, such as the internal hoop stress within the jet or pressure gradients from the external environment, that could further collimate them at even larger radii. The final answer on the collimation of the matter-dominated jets will have to await simulations that extend to larger radii and run for longer times.

\begin{figure}
\begin{center}
\includegraphics[width=0.5\textwidth]{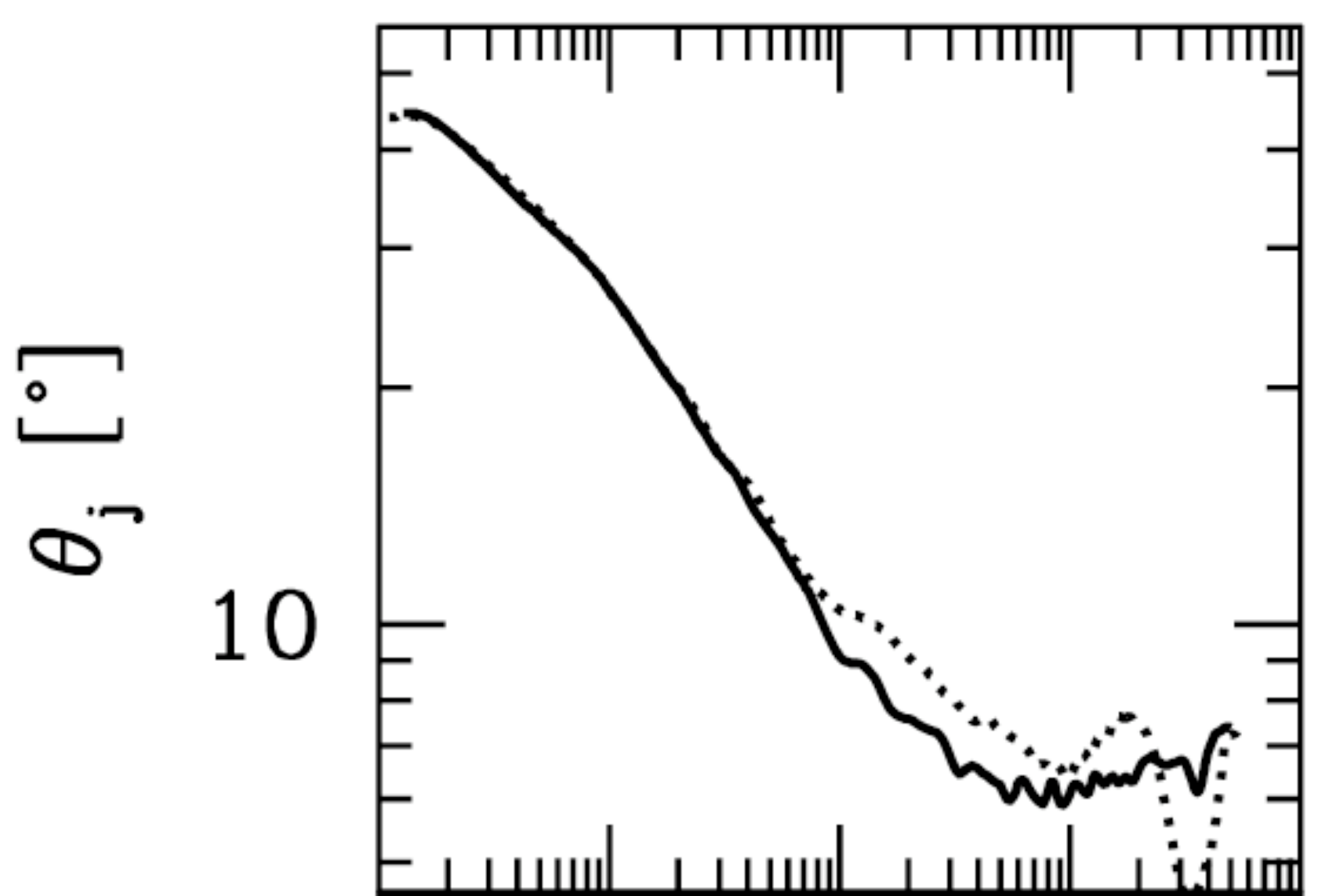}
\end{center}
\caption{Jet opening angle as a function of radius in the simulations of McKinney \cite{mckinney_06}. The horizontal axis is the logarithmically-spaced radius from $1r_G$ to $10^4 r_G$.}
\label{fig7}
\end{figure}

One beneficial attribute of the simulations of McKinney \cite{mckinney_06} is that they covered a much larger spatial extent than previous simulations of this type. The outer boundary of his simulations was at $r_{max} \approx 10^4 r_G$ (as opposed to $r_{max} \approx 100r_G$ in earlier simulations), so he was able to follow the evolution of the jets over nearly 4 orders of magnitude in spatial extent. Again his analysis concentrated on the Poynting-flux dominated jet, which he found starts with only a slightly smaller opening angle than the matter-dominated jet ($\theta \approx 40^\circ$). Like the matter-dominated jet, the Poynting-flux jet is initially confined by the coronal wall. After extending beyond the range of the corona, the Poynting-flux jet is subsequently confined by the surrounding matter-dominated jet. Eventually the Poynting-flux jet is able to collimate itself through the action of internal hoop stresses from the toroidal component of the magnetic field that is carried along in the jet. The self-collimation continues until about $r \sim 500 r_G$, when a final opening angle of $\theta \sim 7^\circ$ is achieved. The gradual collimation of the jet is illustrated in Figure \ref{fig7}.

\section{Caveats}
What I've reviewed up to this point was the state-of-the-art in MHD simulations of jet formation at the time of the 6th Microquasar meeting. I now shift to consider some things we have learned about such simulations since that time.

\subsection{Initial Conditions}
\begin{figure}
\begin{center}
\includegraphics[width=0.6\textwidth]{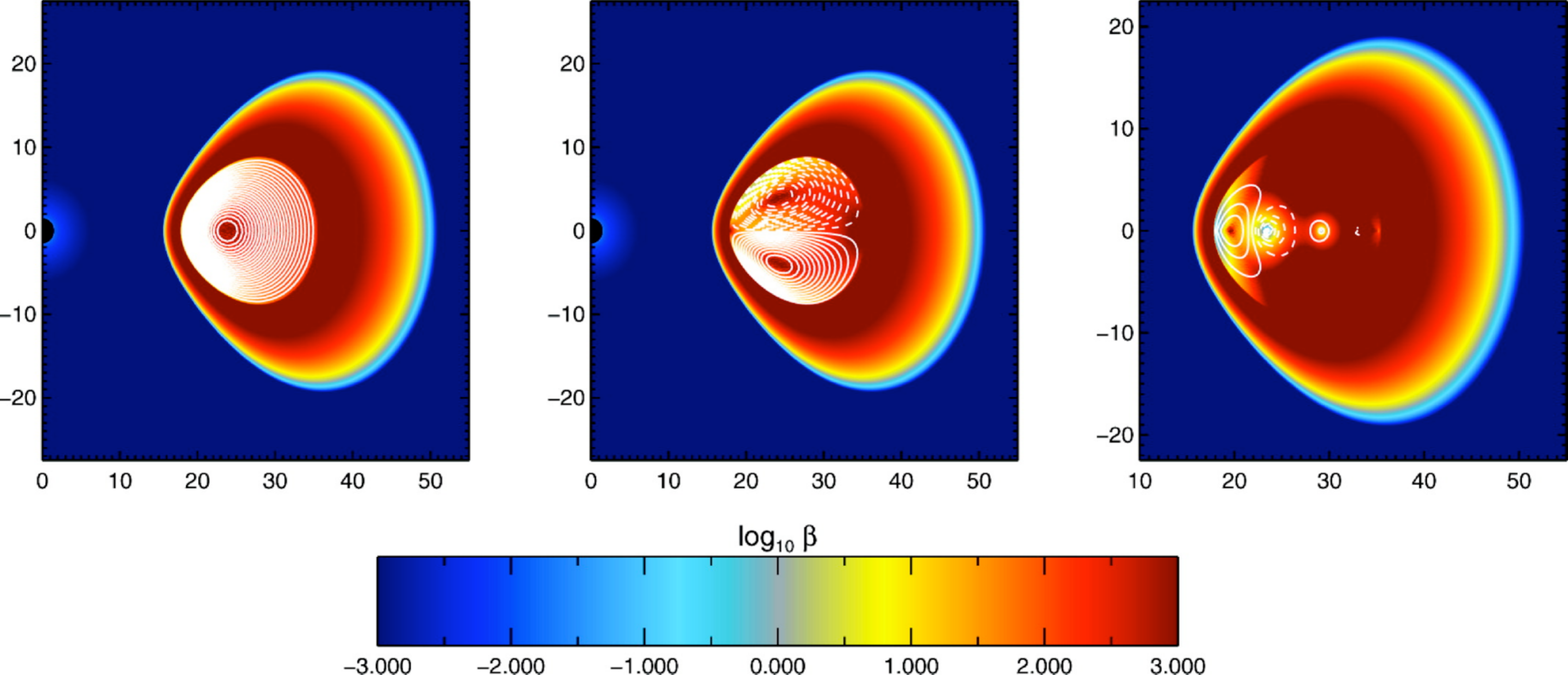}
\end{center}
\caption{Initial configurations of dipole ({\em left panel}), quadrupole ({\em center panel}) and multiple 
loop ({\em right panel}) field topologies. The torus for the multiple loop topology is shown slightly 
zoomed in to illustrate better the initial field structure. White contours denote magnetic field 
lines, color contours the gas $\beta$ parameter. Solid and dashed lines indicate opposite field polarities.}
\label{fig8}
\end{figure}

\begin{figure}
\begin{center}
\includegraphics[width=1.0\textwidth]{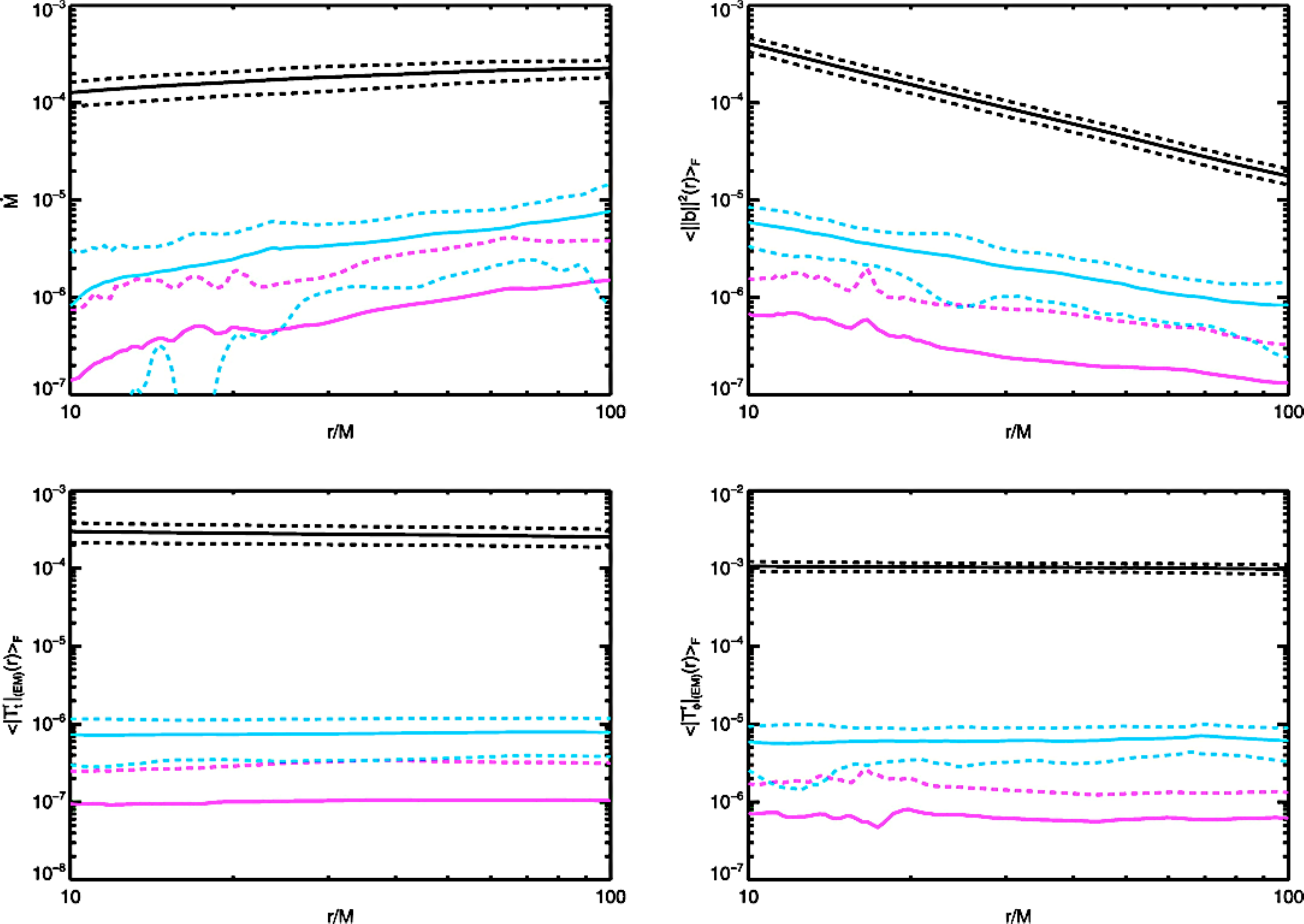}
\end{center}
\caption{Time average of data from unbound outflows as a function of radius for models 
with a dipole ({\em black lines}), quadrupole ({\em blue lines}), and toroidal ({\em magenta lines}) initial magnetic field. These images are taken from \cite{beckwith_08}. Shown are mass outflow rate $\dot{M}$ ({\em top-left panel}), $||b||^2$ ({\em top-right panel}); the electromagnetic energy flux ({\em bottom-left panel}); and the angular momentum flux ({\em bottom-right panel}). Dashed lines show $\pm1$ standard deviation from the average.}
\label{fig9}
\end{figure}

Ideally one would hope that the results of the simulations I have been discussing would be insensitive to their initial conditions, particularly as the simulations are run out to longer times. Since these simulations start from highly artificial initial conditions (a simple gas torus seeded with a simple magnetic field topology) as illustrated in Figure \ref{fig8}, it is particularly important to look for the development of a long-term, steady state disk-jet solution. It is this disk-jet solution that one hopes is independent of the initial conditions. However, because magnetic field in Nature are divergence-free, they can never truly ``forget'' their initial configurations. In analyzing simulations of the type I have been reviewing, Beckwith et al. \cite{beckwith_08} discovered that, although the evolution of the disk is essentially independent of the initial conditions, this can not be said for the jets. By comparing the evolution of initially identical tori seeded with different magnetic field configurations (such as dipole, quadrupole, multiple loops, and toroidal as shown in Figure \ref{fig8}), Beckwith et al. discovered that a critical component in the formation of strong jets is that the magnetic field have a net poloidal flux. Without this, any jets that form appear to be orders of magnitude weaker, as illustrated using a variety of measures in Figure \ref{fig9}.

\section{Future Directions}
I'd like to finish this review by highlighting a couple areas where new contributions are being made in the area of jet formation in MHD simulations. I'm particularly interested in these because they are the current focus of my own research.

\subsection{Tilted Disks and Jets}
A fundamental question regarding jet formation that remains unanswered is: To what characteristics of an accretion disk system can the orientation of a jet be ascribed? Fundamentally there are three possible answers to this question. The first possibility is that the jet orientation is dependent on the angular momentum or spin of the black hole \cite{blandford_77}. This would suggest that the jet should always be perpendicular to the black-hole symmetry plane. The second possibility is that this orientation is dependent on the angular momentum of the accretion disk \cite{lynden-bell_06}, thus orienting the jet perpendicular to its plane of symmetry. A final possibility is that a magnetic field, generated by or near the central object or carried in through the disk, dictates the jet orientation \cite{blandford_82}. In theoretical work it is often assumed that all of these components are aligned, even though there is certainly no requirement that this be the case in real systems. Such an assumption makes it difficult to determine what is setting the orientation of the jet. From an observational perspective, the angular momentum vector of the central black hole is not directly observable. In fact, its orientation has traditionally been inferred from the orientation of the jet and its magnitude from the properties of the disk. This obviously creates a dilemma if one wants to observationally constrain what is setting the orientation of a jet.

However, direct MHD numerical simulations of tilted black-hole accretion disks, such as those presented in \cite{fragile_07}, allow one to explore situations in which the black-hole spin, disk angular momentum, and magnetic field are not all aligned. One then need only observe the orientation of any resulting jets in these simulations to address the question of what sets their orientation. The disks in tilted simulations have also been shown to precess \cite{fragile_07} \footnote{Movies of the tilted disk simulations of Fragile can be seen at http://fragilep.people.cofc.edu/research/tilted.html .}, so another interesting question is whether or not jets can be forced to precess by such a disk.

\subsection{Beyond Ideal MHD}
\begin{figure}
\begin{center}
\includegraphics[width=0.8\textwidth]{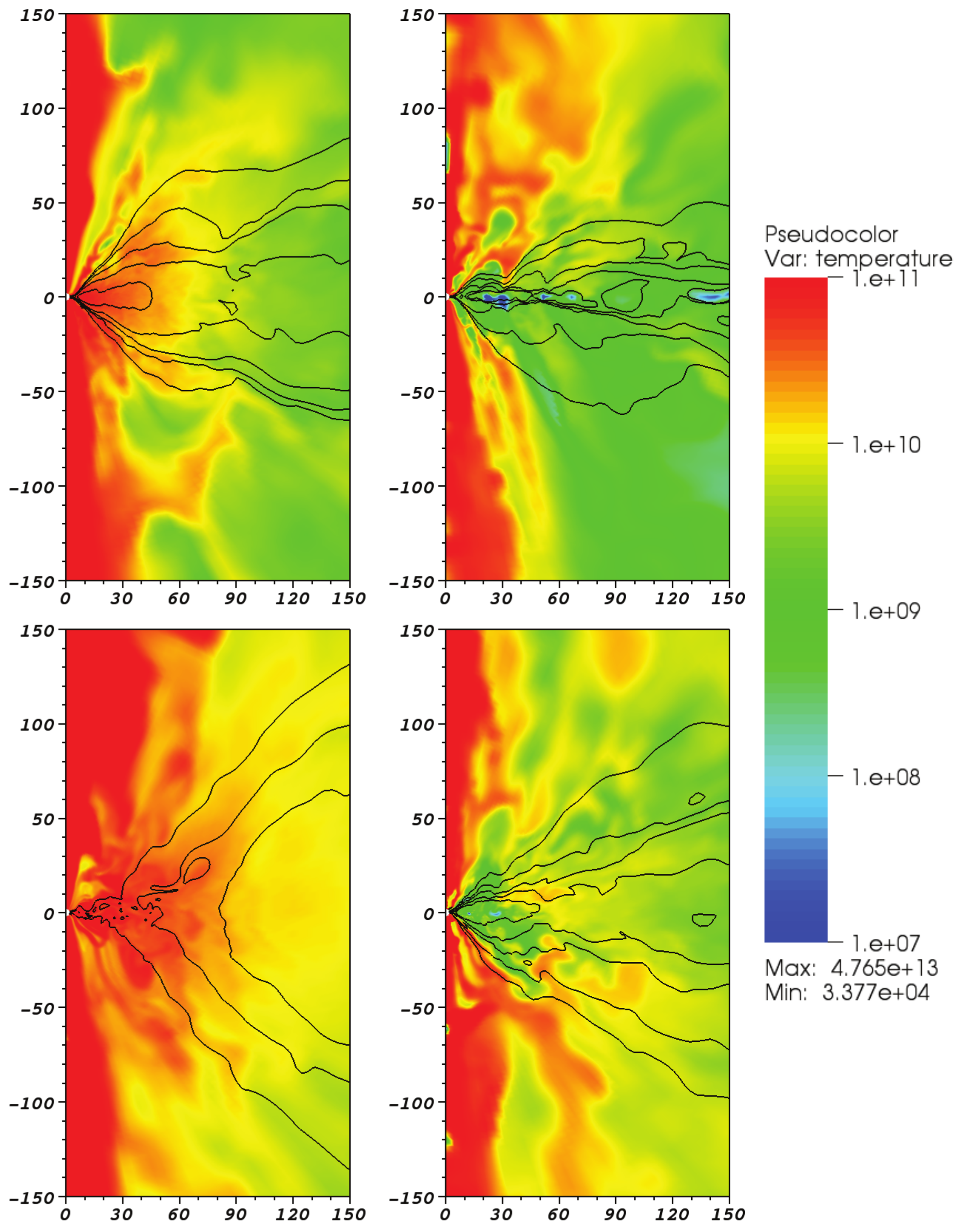}
\end{center}
\caption{Pseudo-color plots of the logarithm of temperature with contours of the logarithm of density. The {\em upper-left} panel 
is from an internal energy model (like De Villiers \& Hawley); the {\em upper-right} panel
is from an internal energy plus cooling model; the {\em lower-left} panel is from a total energy model (like McKinney \& Gammie); and the {\em lower-right} panel is from a total energy plus cooling model from Fragile \& Meier \cite{fragile_08}.}
\label{fig10}
\end{figure}

\begin{figure}
\begin{center}
\includegraphics[width=0.8\textwidth]{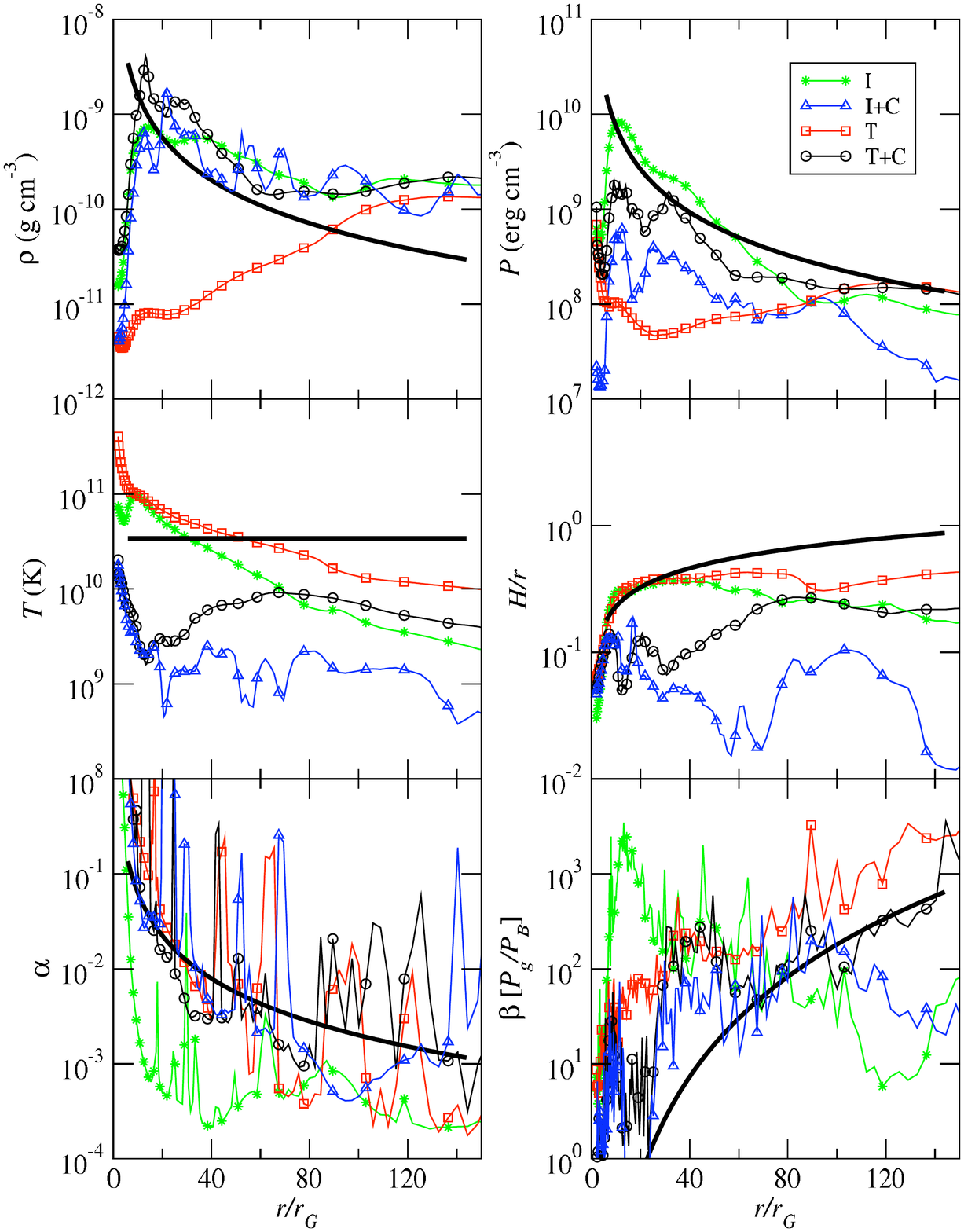}
\end{center}
\caption{Main disk properties plotted as a function of radius for the internal-energy model I, the internal-energy plus cooling model I+C, the total-energy model T, and the total-energy plus cooling model T+C from Fragile \& Meier \cite{fragile_08}. Plotted variables are the density $\rho$, gas pressure $P$, gas temperature $T$, temperature scale-height $H/r$, stress parameter $\alpha$, and the equipartition parameter $\beta$.}
\label{fig11}
\end{figure}

Perhaps the greatest shortcoming of the numerical simulations of black-hole accretion disks and jets carried out up until this past year, has been the very incomplete treatment of thermodynamics. This is also the area in which the simulations of McKinney and Hawley differ the most. Hawley and his collaborators use a formulation of the MHD equations that evolves the internal (not total) energy of the gas \cite{devilliers_03}. Because their code does not conserve total energy and does not include any terms to capture dissipative heating processes, any heat created by disk turbulence is simply lost from the simulation. In a sense, the entire flow is artificially cooled, although heating from compression and shocks is still treated. McKinney, on the other hand, uses a total-energy conserving formulation \cite{gammie_03}, so that any kinetic and magnetic energy dissipated within the disk is automatically recaptured as heat. However, McKinney's code does not include any radiative cooling processes. As such, it strictly only applies to radiatively inefficient flows such as the disk around Sgr A*. Recently, Fragile \& Meier \cite{fragile_08} performed the first global numerical simulations of black-hole accretion disks that included {\em both} a total-energy formulation to capture dissipative heating processes {\em and} a radiative cooling function that approximated cooling due to bremsstrahlung, synchrotron, and inverse-Compton processes. Figures \ref{fig10} and \ref{fig11} show how these different numerical approaches affect the final state of the disk. The key result is that the radiatively-cooled disk is much cooler and much closer to equipartition than the other models. The approach toward equipartition may signal a transition to a magnetically-dominated accretion flow (MDAF) as proposed by Meier \cite{meier_05}.

The other important point about including real radiative cooling processes in numerical simulations is that it opens the door to the possibility of making real connections between simulations and observations. This is an exciting new possibility!

\section{Conclusions}
The main take-away points from this review are

\begin{itemize}
\item{Jets are consistently formed in MHD simulations. They
do \emph{not} require pre-existing large-scale magnetic fields. They
do, however, require a net poloidal flux.}
\item{The jets in these numerical simulations have a 2-component, sheath configuration, with a 
slower, matter-dominated jet on the outside and a 
faster, magnetically-dominated jet on the inside.}
\item{The era of radiatively-cooled MHD simulations of black-hole accretion disks is beginning. This will 
allow more direct comparison between numerical simulations and observations. One desired outcome of this work is an understanding of state transitions in microquasars, which may be related to changes in the thermodynamic state of the disk.}
\end{itemize}

\end{document}